\documentclass[preprint,showpacs,amssymb,amsmath,aps,pra]{revtex4-1}
\usepackage{graphics}
\usepackage{bm}
\usepackage[spanish,english]{babel}
\usepackage{graphics}
\usepackage{bm}
\usepackage{amsmath}
\usepackage{amssymb}
\begin{document}

\title{Discord and information deficit in the $XX$ chain}
\author{L. Ciliberti, N. Canosa, R. Rossignoli}
\affiliation{Departamento de F\'{\i}sica-IFLP,
Universidad Nacional de La Plata, C.C. 67, La Plata (1900), Argentina}
\begin{abstract}
We examine the quantum correlations of spin pairs in the cyclic $XX$ spin $1/2$
chain in a transverse field, through the analysis of the quantum discord, the
geometric discord and the information deficit. It is shown that while these
quantities provide the same qualitative information, being non-zero for all
temperatures and separations and exhibiting the same type of asymptotic
behavior for large temperatures or separations, important differences arise in
the minimizing local measurement that defines them. Whereas the quantum discord
prefers a spin measurement perpendicular to the transverse field, the geometric
discord and information deficit exhibit a perpendicular to parallel transition
as the field increases, which subsists at all temperatures and for all
separations. Moreover, it is shown that such transition signals the change from
a Bell state to an aligned separable state of the dominant eigenstate of the reduced
density matrix of the pair. Full exact results for both the thermodynamic limit
and the finite chain are provided, through the Jordan-Wigner fermionization.
\end{abstract}
\maketitle

\section{Introduction}

The investigation of quantum correlations in mixed states is presently
attracting strong attention \cite{Mo.12}. While in bipartite pure states such
correlations can be identified with entanglement, it was recently recognized
that separable (non-entangled) bipartite mixed states, defined as states which
can be created by local operations and classical communication, and which are
therefore convex mixtures of product states \cite{RW.89}, may still exhibit
useful quantum correlations, stemming from the non-commutativity  of the
different products. The mixed state based quantum algorithm introduced by Knill
and Laflamme (KL) \cite{KL.98} has shown that an exponential speed-up over
classical algorithms can in fact be achieved without entanglement
\cite{DFC.05}, in contrast with the case of pure states \cite{JL.03}.

This has turned the attention to alternative measures of quantum correlations
for mixed states, such as the quantum discord \cite{OZ.01,HV.01,Zu.03,Mo.12},
which are able to capture the quantumness of such mixed states, vanishing just
for states diagonal in a product basis and coinciding with entanglement in the
pure state limit. A finite discord between the control qubit and the remaining
maximally mixed qubits was in fact found in the KL algorithm \cite{DSC.08},
renewing the interest on this measure \cite{Dat.09,SL.09,Ve.10,FA.10,FCOC.11}.
Other measures with similar properties include the closely related one-way
information deficit \cite{HHH.05,SKB.11,Mo.12}, the geometric discord
\cite{DVB.10}, which allows an easier evaluation, and the generalized entropic
measures of ref.\ \cite{RCC.10}, which include the previous ones as particular
cases. Various applications and operational interpretations of the quantum
discord and related measures were recently provided
\cite{Mo.12,Dat.09,SKB.11,MD.11,CC.11,PG.11,GA.12,TG.12,BLC.12}. We remark that
all these measures require a minimization over local measurements in one of the
constituents (which can be viewed as the determination of the least disturbing
local measurement \cite{RCC.11}), which makes their evaluation difficult in
systems with high dimensionality.

Spin chains provide an interesting scenario for studying these measures and
their relation with criticality
\cite{Dd.08,SA.09,MG.10,WR.10,CRC.10,WR.11,BQL.11,YH.11,CCR.12,Mo.12,T.11,SC.13}.
In particular, the state of a spin pair in an entangled ground state (GS) is in
general a mixed state, entailing that differences between the entanglement and
the quantum discord of a spin pair will arise already at zero temperature
\cite{Dd.08,MG.10,CRC.10}. These differences become significant in the exact
ground states of finite XY chains for transverse fields lower than the critical
field $B_c$ \cite{CRC.10}, with the quantum discord reaching full range in this
region.

The aim of the present work is to analyze in detail the behavior of the quantum
discord, the geometric discord and the one-way information deficit of spin
pairs in chains with $XX$-type first neighbor couplings in a transverse field,
at both zero and finite temperature. Such model is particularly interesting for
both quantum information and condensed matter physics, exhibiting distinct
features such as eigenstates with definite magnetization along the field axis
and a special critical behavior \cite{SS.99}. It is first shown that in
contrast with entanglement \cite{Wa.02,CR.07,FH.07}, discord-type measures
exhibit common features such as a non-zero value for all separations $L$ at all
temperatures $T>0$. Exact asymptotic expressions for the decay with $L$ and $T$
will be provided, on the basis of the exact treatment based on the
Jordan-Wigner fermionization \cite{CR.07,FH.07,LSM.61,PF.09}. Nonetheless, we
will also show that important differences between the quantum discord on the
one side, and the geometric discord and information deficit on the other side,
do arise in the minimizing local spin measurement. While in the quantum discord
the latter is always orthogonal to the transverse field (even at strong fields
if $T>0$), in the geometric discord and information deficit it exhibits a
perpendicular to parallel transition  as the field increases, at a field lower
than the $T=0$ critical field $B_c$. Such transition in the minimizing
measurement is present at all temperatures and separations, and as will be
shown, is a signature of the transition from a Bell state to a separable
aligned state of the dominant eigenstate of the reduced density matrix of the
pair. This difference indicates a distinct response of the minimizing
measurement in these quantities to the onset of quantum correlations.

In Section II we summarize the main features of the previous  measures,
including the equations that determine the stationary local measurements. The
application to the spin $1/2$ $XX$ chain is made in section III, where we first
discuss some general properties of these measures in this model and show that
spin measurements parallel and perpendicular to the field are always
stationary. We then consider in detail the thermodynamic limit and the finite
case. Details of the exact calculation are provided in the Appendix.
Conclusions are finally given in IV.

\section{Discord and generalized information deficit}

Let us consider a bipartite quantum  system $A+B$ initially in a state
$\rho_{AB}$. A local complete projective measurement $M_B$ on system $B$ is
defined by a set of orthogonal projectors $\Pi_j^B=I_A\otimes\Pi_j$, where
$\Pi_j=|j_B\rangle\langle j_B|$ are one dimensional projectors satisfying
$\sum_j \Pi_j=I_B$,  $\Pi_j\Pi_k=\delta_{jk}\Pi_k$. The state of the total
system after an unread measurement of this type becomes
\begin{equation}
\rho'_{AB}=\sum_j \Pi_j^B\rho_{AB}\Pi_j^B\,.\label{rhop}
\end{equation}
In \cite{RCC.10,RCC.11} we considered the {\it minimum generalized
information loss} by such measurement,
 \begin{equation}
 I^B_f=\mathop{\rm Min}_{M_B} S_f(\rho'_{AB})-S_f(\rho_{AB})\,,
\label{If}
\end{equation}
where $S_f(\rho)={\rm Tr}\, f(\rho)$ denotes a general entropic form, with $f$
a smooth strictly concave function for $p\in[0,1]$, satisfying $f(0)=f(1)=0$
\cite{CR.02}. Eq.\ (\ref{If}) satisfies $I_f^B\geq 0$ for any such $f$,
becoming  the generalized entanglement entropy $S_f(\rho_B)=S_f(\rho_A)$ in the
case of pure states ($\rho^2_{AB}=\rho_{AB}$). However, it can be non-zero in
separable {\it mixed} states, vanishing just for states which are already of
the form (\ref{rhop}) \cite{RCC.10}, i.e., states which remain unchanged after
the local measurement $M_B$ and are hence diagonal in a product basis
$\{|{i_j}_A\rangle\otimes|j_B\rangle\}$.  The positivity of $I_f^B$ $\forall$
$S_f$ follows from the majorization relation  $\rho'_{AB}\prec\rho_{AB}$
satisfied by (\ref{rhop}) \cite{RCC.10,RCC.11,Ww.78}.

If $f(\rho)=-\rho\log_2\rho$, $S_f(\rho)$ becomes the von Neumann entropy
$S(\rho)$ and Eq.\ (\ref{If}) becomes the {\it one way information deficit}
\cite{HHH.05,SKB.11,Mo.12}, which we will denote as $I_1^B$. It can be
rewritten in terms of the relative entropy \cite{Ww.78,Ve.02} $S(\rho
||\rho')=-{\rm Tr}\,\rho (\log_2\rho'-\log_2\rho)$ as \cite{RCC.10}
\begin{equation}
 I^B_1=\mathop{\rm Min}_{M_B} S(\rho'_{AB})-S(\rho_{AB})\,
 =\mathop{\rm Min}_{M_B}S(\rho_{AB} ||\rho'_{AB})\,.
\label{ID}
\end{equation}
For a pure state, $I_1^B$ becomes the standard entanglement entropy
$S(\rho_A)=S(\rho_B)$.

If $f(\rho)=2\rho(1-\rho)$, $S_f(\rho)$ becomes the so called {\it
linear entropy} $S_2(\rho)=2(1-{\rm Tr}\,\rho^2)$ and Eq.\ (\ref{If}) becomes
\begin{equation}
I_2^B = 2\mathop{\rm Min}_{M_B}{\rm Tr}\, (\rho^{2}_{AB}-{\rho'}_{AB}^{\,2})=
2\mathop{\rm Min}_{\rho'_{AB}}||\rho_{AB}
-\rho'_{AB}||^2\,, \label{IB2}
 \end{equation}
where $||O||^2={\rm Tr}\,O^\dagger O$ and the last minimization can be
extended to any state of the general form (\ref{rhop}). It is then seen that
(\ref{IB2}) is proportional to the {\it geometric discord} introduced in
\cite{DVB.10}, defined as the square of the minimum Hilbert-Schmidt distance
from $\rho_{AB}$ to a classically correlated state with respect to $B$. For
pure states $I_2^B$ becomes the squared concurrence $C^2_{AB}$ \cite{HW.97},
which for such states is just the linear entropy of any of the subsystems
\cite{Ca.03}.

Both measures (\ref{ID})--(\ref{IB2}) can then be regarded as particular cases
of the generalized information deficit (\ref{If}). We may similarly define
\cite{RCC.11} $I_q^B=S_q(\rho'_{AB})-S_q(\rho_{AB})$ for entropies $S_q(\rho)$
associated to $f(\rho)=(\rho-\rho^q)/(1-2^{1-q})$, $q>0$ \cite{TS.09}.
$I_q^B$ reduces to
(\ref{IB2}) for $q=2$ and to (\ref{ID}) for $q\rightarrow 1$
($S_q(\rho)\rightarrow S(\rho)$ in this limit). Normalization of $f(\rho)$ was
chosen such that $I_f^B=1$  $\forall\, S_f$ for a two-qubit Bell state

On the other hand, the quantum discord \cite{OZ.01,HV.01} for a
measurement in $B$ can be written as
\begin{equation}
D^B =\mathop{\rm Min}_{M_B}S(A|M_B)-S(A|B)=
\mathop{\rm Min}_{M_B}[I_1^{M_B}(\rho_{AB})-I_1^{M_B}(\rho_B)]\,, \label{DI}
 \end{equation}
where $S(A|M_B)$ denotes the conditional von Neumann entropy of $A$ given a
measurement $M_B$ in $B$,  $S(A|B)=S(\rho_{AB})-S(\rho_B)$ is the quantum
conditional entropy and the last expression is the result for a complete
projective measurement $M_B$, which is the case we will here consider. $D^B$ is
just the minimum decrease of the {\it mutual information} $S(A)-S(A|B)$ after an
unread measurement in $B$ \cite{OZ.01,HV.01}. We then have $D^B\leq I_1^B$,
with $D^B=I_1^B$ when the minimizing measurements for $D^B$ and $I_1^B$
coincide and $\rho'_B=\rho_B$. Nonetheless, like $I_1^B$, $D^B\geq 0$,
vanishing just for the classically correlated states (\ref{rhop}) and reducing
to the entanglement entropy $S(\rho_A)=S(\rho_B)$ for pure states $\rho_{AB}$.

However, important differences between $I_1^B$ (or in general $I_f^B$) and
$D^B$ may arise in the minimizing measurement. While for a general classically
correlated state of the form (\ref{rhop}) the minimum (0) for {\it both} $D^B$
and {\it all} $I_f^B$ is attained for a measurement in the local basis defined
by the projectors $\Pi_j^B$ (i.e., the pointer basis \cite{OZ.01,HV.01}), in
the particular case of product states $\rho_A\otimes \rho_B$, $D^B$ (but not
$I_f^B$) becomes the same for {\it any} $M_B$, as for such states
$S(A|M_B)=S(A)$ $\forall$ $M_B$. The same holds for pure states $\rho_{AB}$,
where $D^B$ is again the same for any $M_B$, as here $S(A|M_B)=0$ $\forall$
$M_B$ of the present form, whereas the minimum $I_f^B$ is obtained, for {\it any} $S_f$, for a
measurement $M_B$ in the local part of the Schmidt basis \cite{RCC.10}, i.e.,
again in the basis of eigenstates of the reduced state $\rho_B$. These
differences will have important consequences for the results of the next
section, leading to a quite different response of the minimizing measurement to
the onset of quantum correlations. They reflect the fact that while in $I_f^B$
one is looking for the least disturbing local measurement, such  that
$\rho'_{AB}$ is as close as possible to $\rho_{AB}$, in $D^B$ the search is for
the measurement in $B$ which makes the ensuing conditional entropy smallest,
i.e., by which one can learn the most about $A$.

We also remark that the determination of the minimizing measurement $M_B$ is in
general a difficult problem. Complete projective measurements at $B$ are
determined by $d^2_B-d_B$ real parameters if $B$ has a Hilbert space of
dimension $d_B$, growing then exponentially with the number of components of
$B$. For $I_f^B$, the minimizing measurement should fulfill the stationary
condition \cite{RCC.11,RMC.12}
\begin{equation}
{\rm Tr}_{A}[f'(\rho'_{AB}),\rho_{AB}]=0 \,,
\label{stat}
\end{equation}
which leads to $d_B(d_B-1)$ real equations
\cite{RCC.11}.  In the quantum discord (\ref{DI}), an additional
term $-[f'(\rho'_B),\rho_B]$ is to be added in (\ref{stat}), with
$f(\rho)=-\rho\log_2\rho$ \cite{RCC.11} (see also \cite{Ar.10,AD.11}).

Nevertheless, in the case of the geometric discord $I_2$, the final equations
can be simplified considerably. In particular, for a general mixed state of two
qubits
\begin{equation}\rho_{AB}
={\textstyle\frac{1}{4}}(I+\bm{r}_A\cdot\bm{\sigma}_A+\bm{r}_B \cdot
 \bm{\sigma}_B +\bm{\sigma}^t_A J\bm{\sigma}_B)\,,\label{rq}\end{equation}
where $\bm{\sigma}=2\bm{s}$ are the Pauli matrices,
$\bm{\sigma}_A=\bm{\sigma}\otimes I$, $\bm{\sigma}_B=I\otimes\bm{\sigma}$, 
$\langle\bm{\sigma}_{A,B}\rangle=\bm{r}_{A,B}$ and
$J=\langle \bm{\sigma}_A\bm{\sigma}_B^t\rangle$, it can be shown that
\cite{DVB.10}
\begin{equation}
I_2^B={\textstyle\frac{1}{2}}({\rm tr}\,M_2-\lambda_1)\,,\label{IB2x}
\end{equation}
where $\lambda_1$ is the largest eigenvalue of the positive semi-definite
$3\times 3$ matrix $M_2=\bm{r}_B\bm{r}_B^t+J^tJ$. The minimizing $M_B$ is a
spin measurement along the direction of the associated eigenvector $\bm{k}_1$
of $M_2$. A closed expression for $I_3^B$ can also be obtained for this case
\cite{RCC.11}.

\section{Application to the XX model}
We now consider a chain of $N$ spins $\bm{s}_{i}$ with first neighbor $XX$
couplings in a uniform transverse magnetic field. The Hamiltonian reads
\begin{equation}
H=\sum_{i}\,B s_{iz}-
J(s_{ix}s_{i+1,x}+s_{iy}s_{i+1,y})\,, \label{H1}
\end{equation}
and is obviously invariant under rotations around the $z$ axis, satisfying
$[H,S_z]=0$, with $S_z=\sum_i s_{iz}$ the $z$-component of total spin.  Its
eigenstates can then be characterized by the total magnetization $M$ along $z$.
The sign of the field $B$ and the coupling strength $J$ can be changed by local
rotations $e^{i\pi s_{jz}}$ at all and even spins $j$ respectively (assuming
$N$ even in the cyclic case $N+1\equiv 1$), so that we will set in what follows
$B\geq 0$, $J\geq 0$.

We will examine the spin $1/2$ case, where exact results for finite $N$ as well
as the thermodynamic limit $N\rightarrow\infty$ can be obtained via the
Jordan-Wigner fermionization (see Appendix). We will focus on the cyclic case
$N+1\equiv 1$, where pair correlations between spins $i$ and $j$ in the ground
state or in the thermal state $\rho\propto\exp[-\beta H]$ will depend just on
the separation $L=|i-j|$.

For any global state $\rho$ satisfying $[\rho,S_z]=0$, the reduced state
$\rho_{ij}={\rm Tr}_{\overline{ij}}\,\rho$ of any pair $i\neq j$ will commute
with $s_{iz}+s_{jz}$. In the cyclic case, $\rho_L\equiv\rho_{ij}$ will then
have the form
  \begin{eqnarray}
\rho_{L}&=&\left(\begin{array}{cccc}p^+_L&0&0&0\\0&p_L&\alpha_L&0\\
 0&\alpha_L&p_L&0\\0&0&0&p^-_L\end{array}\right)\label{rij}\\
 &=&p^+_L|\!\uparrow\uparrow\rangle\langle\uparrow\uparrow\!|+
p^-_L|\!\downarrow\downarrow\rangle\langle\downarrow\downarrow\!|\nonumber\\&&
+(p_L+\alpha_L)|\Psi_+\rangle\langle\Psi_+|+
 (p_L-\alpha_L)|\Psi_-\rangle\langle\Psi_-|\,,\label{rijxx}\end{eqnarray}
where (\ref{rij}) is the representation in the standard basis and (\ref{rijxx})
the eigenvector expansion, with
$|\Psi_{\pm}\rangle=\frac{|\uparrow\downarrow\rangle
\pm|\downarrow\uparrow\rangle}
{\sqrt{2}}$ Bell states. Here $p^+_L+p^-_L+2p_L=1$, with
 \begin{eqnarray}
 p^\pm_L&=&{\textstyle\frac{1}{4}}\pm\langle s_{z}\rangle+\langle
 s_{iz}s_{jz}\rangle \,,\\
 \alpha_L&=&\langle s_{ix}s_{jx}+s_{iy}s_{jy}\rangle\,,
 \label{vm}
 \end{eqnarray}
and $\langle s_z\rangle=\langle S_z\rangle/N$ the intensive average
magnetization along $z$. It corresponds to $\bm{r}_A=\bm{r}_B=(0,0,2\langle
s_z\rangle)$ and $J_{\mu\nu}=\delta_{\mu\nu}J_\mu$ in (\ref{rq}), with
$2\langle s_z\rangle=p_L^+-p_L^-$, $J_x=J_y=2\alpha_L$, $J_z=1-4p_L$.

The eigenvectors of $\rho_{L}$ in the ground or thermal state will not depend
then on the field or separation. For $B\geq 0$ and $J\geq 0$ in (\ref{H1}),
$p^-_L\geq p^+_L$
and $\alpha_L\geq0$. The largest eigenvalue of $\rho_L$ will then correspond to
the Bell state $|\Psi_+\rangle$ if
\begin{equation}\alpha_L>\alpha_L^c=p^-_L-p_L\,,\label{critl}\end{equation}
and to the aligned separable state $|\!\!\downarrow\downarrow\rangle$ if
$\alpha_L<\alpha^c_L$. Hence, in the ground state we may expect as the field
decreases a transition from  $|\!\!\downarrow\downarrow\rangle$ to
$|\Psi_+\rangle$ in the dominant eigenstate of $\rho_{L}$, at a certain field
$B^L_c\leq B_c$, where $B_c=J$ denotes the $T=0$ critical field (such that the
ground state is fully aligned ($M=-N/2$) for $B>B_c$). We will see such
crossing reflected in the transition exhibited by the geometric discord and the
information deficit (but not the quantum discord). We will also find the same
effect at finite temperatures.

\subsection{Parallel and perpendicular geometric discord and information deficit}
We first discuss the general properties of the discord and information deficit
of the states (\ref{rij}). Due to the permutation symmetry of $\rho_{ij}$, we
will omit in what follows the superscript $B$ (i.e., $j$) in $I_f$ and $D$, as
$I_f^B=I_f^A$, $D^B=D^A$. For $\alpha_L=0$, $\rho_{L}$ is diagonal in the
standard basis and  will then have zero entanglement and discord: $E=D=I_f=0$ 
$\forall$ $S_f$. It will be, however, classically correlated, being a product
state $\rho_i\otimes\rho_j$ only when  $p_L=\sqrt{p^+_Lp^-_L}$ (in which case
$\rho_i=\rho_j=\sqrt{p^+_L}|\!\uparrow\rangle\langle\uparrow\!|+
\sqrt{p^-_L}|\!\downarrow\rangle\langle\downarrow\!|$).

Quantum correlations will then be driven solely by $\alpha_L$, and will lead to
a finite value of $D$ and $I_f$ $\forall$ $\alpha_L\neq 0$. The geometric
discord (\ref{IB2}) for such state can be evaluated immediately with Eq.\
(\ref{IB2x}) (here $(M_2)_{\mu\nu}=\delta_{\mu\nu} \lambda_\mu$,  with
$\lambda_x=\lambda_y=J_x^2$, $\lambda_z=J_z^2+|\bm{r}_B|^2$) and reads
\begin{equation}
I_2=\left\{\begin{array}{ll}I_2^z=4\alpha_L^2\,,&|\alpha_L|\leq\alpha_L^t\\
\, I_2^{\perp}=2(\alpha_L^2+{\alpha_L^t}^2)\,,&|\alpha_L|\geq \alpha_L^t\,,
 \end{array}\right.\label{I2x}\end{equation}
where $\alpha_L^t=\frac{\sqrt{\lambda_z}}{2}=
\sqrt{\frac{(p_L^--p_L)^2+(p_L-p_L^+)^2}{2}}$ and the superscript in $I_2$
indicates the direction of the minimizing local spin measurement (along $z$ if
$|\alpha_L|<\alpha_L^t$ and along any orthogonal direction $\bm{k}$  if
$|\alpha_L|>\alpha^t_L$). Hence, $I_2$ increases first quadratically with
$\alpha_L$ and exhibits then a parallel $\rightarrow$ perpendicular transition
at $\alpha_L=\alpha_L^t$, corresponding to  a transition field $B_t^L$. For
$p^-_L>p_L$ such transition {\it correlates with that exhibited by the dominant
eigenstate} of $\rho_{L}$ (Eq.\ (\ref{critl})). In fact, if
$|p^+_L-p_L|=|p^-_L-p_L|$ and $p^-_L>p_L$, $\alpha_L^t=\alpha_L^c$.

Eq.\ (\ref{I2x})  is to be contrasted with the concurrence of $\rho_L$, which
requires a finite threshold value of $\alpha_L$:
 \begin{equation}
 C=2{\rm Max}[|\alpha_L|-\sqrt{p_L^+ p_L^-},0]\,.
 \label{Cij}
 \end{equation}
Hence, discord-type quantum correlations with zero entanglement will arise for
$0<|\alpha_L|\leq \sqrt{p^+_Lp^-_L}$.

The behavior of the generalized information deficit (\ref{If}) is similar to
that of the geometric discord. For a spin measurement along a vector $\bm{k}$
forming an angle $\gamma$ with the $z$ axis, the eigenvalues of the
post-measurement state $\rho'_{L}$ are, setting $\delta=\langle
s_z\rangle=(p_L^+-p_L^-)/2$ and $\mu,\nu=\pm 1$,
\[{p'}_\mu^\nu={\textstyle\frac{1+2\nu\delta\cos\gamma+
\mu\sqrt{[(1-4p_L)\cos\gamma+2\nu
 \delta]^2+4\alpha_L^2\sin^2\gamma}}{4}}\,.\]
It is then verified that $\partial I_f^\gamma/\partial\gamma=0$ at $\gamma=0$
and $\gamma=\pi/2$: {\it Both parallel ($\gamma=0$) and perpendicular
($\gamma=\pi/2$) measurements are always stationary}, in agreement with the
general considerations of \cite{RCC.11}. Intermediate minima may also arise for
a general $S_f$,  but the essential competition is between $I_f^z\equiv I_f^0$
and $I_f^{\perp}\equiv I_f^{\pi/2}$.

For small $\alpha_L$ and $\delta\neq 0$, the minimum $I_f^\gamma$ for any
 $S_f$ will be obtained for $\gamma=0$, with
 \begin{eqnarray} I_f^z&=&2f(p_L)-f(p_L+\alpha_L)
 -f(p_L-\alpha_L)\approx k_f\alpha_L^2\,,\label{Iza}\end{eqnarray}
where $k_f=|f''(p_L)|$ (we assumed here $p_L\neq 0$). Hence, as $\alpha_L$
increases from $0$, all $I_f$  will exhibit an initial {\it quadratic increase}
with $\alpha_L$, like the geometric discord.

On the other hand, if $\delta=0$ ($p_L^+=p_L^-$), as in the case of zero field
in the ground or thermal state, the minimum $I_f^\gamma$ for {\it any} $S_f$ is
attained for $\gamma=0$ if  $|\alpha_L|<\alpha_L^t$ and for $\gamma=\pi/2$ if
$|\alpha_L|>\alpha_L^t$, where $\alpha_L^t=|\frac{1}{2}-2p_L|=|p^-_L-p_L|$ as
in Eq.\ (\ref{I2x}).  Hence, {\it all } $I_f$'s will in this case exhibit, like
the geometric discord, a parallel $\rightarrow$ perpendicular transition at the
{\it same} value of $\alpha_L$.  Moreover, for $p^-_L>p_L$, $\alpha_L^t$
coincides in this case {\it exactly} with $\alpha_L^c$, i.e, with the value
where the dominant eigenstate of $\rho_{L}$ becomes a Bell state.

The same behavior occurs  when $p^\pm_L=\frac{1}{4}\pm \delta$ (implying
$p_L=\frac{1}{4}$)  with $\alpha_L$, $\delta$ small, a typical situation to be
encountered at high temperatures or large separations. A series expansion of
$I_f^\gamma$ leads to $I_f^\gamma\approx
k_f[\alpha_L^2-\frac{1}{2}\sin^2\gamma(\alpha_L^2-\delta^2)]$, where
$k_f=|f''(1/4)|$, implying again
\begin{equation}
I_f=\left\{\begin{array}{ll}I_f^z\approx k_f\,
\alpha_L^2\,,&\;|\alpha_L|<|\delta|\\
I_f^\perp\approx k_f(\alpha_L^2+\delta^2)/2\,,&\;|\alpha_L|>|\delta|
\end{array}\right.\,,\label{grg}\end{equation}
with $\alpha_L^t=|\delta|=\alpha_L^c$ if $p_L^->p_L$. Hence we obtain in this
case a universal parallel $\rightarrow$ transverse transition at
$|\alpha_L|=|\delta|$ $\forall$ $S_f$ and $L$. In other words, all $I_f$ behave
like the geometric discord in this limit.

In contrast, the minimizing projective spin measurement of the quantum discord $D$
will not exhibit such transition for the present Hamiltonian. We obtain, setting
now $f(p)=-p\,\log_2 p$,
\begin{equation}
D^\gamma=I_1^\gamma-\sum_{\nu=\pm}[f({\textstyle\frac{1}{2}}+
\nu\delta\cos\gamma)-f({\textstyle\frac{1}{2}}+\nu\delta)]\,.
 \label{De}\end{equation}
Hence, $D^z\equiv D^0=I_1^z$,  but $D^\gamma<I_1^\gamma$ if  $|\cos\gamma|<1$
and $\delta\neq 0$ (however, at zero field, $\delta=0$ and
$D^\gamma=I_1^\gamma$ $\forall$ $\gamma$, implying $D=I_1$). While both
$\gamma=0$ and $\gamma=\pi/2$ are again always stationary, the minimum
$D^\gamma$ will be always obtained for $\gamma=\pi/2$ ($D=D^{\perp}$) for the
actual reduced states derived from the ground or thermal state determined by
$H$, directly reflecting  the spin-spin coupling in (\ref{H1}) (which involves
the spin components perpendicular to the field axis). This will also occur for
small $\alpha_L$, since in this limit the actual values of $p_L^\pm$ will
correspond to a product state, entailing no preferred direction in $D^\gamma$
for $\alpha_L=0$. In fact, for small $\alpha_L$ and $\gamma=\pi/2$, Eq.\
(\ref{De}) leads, for $p_L=\sqrt{p^+_Lp^-_L}>0$, to
\begin{equation}
D^{\perp}\approx {\textstyle\frac{1}{\ln 2}}({\textstyle \frac{1}{p_L}
-\frac{{\rm arctanh}\, 2\delta}{\delta}})\,\alpha_L^2\,,\label{Dperpa}
\end{equation}
which is always smaller than $D^z=I_f^z\approx \frac{\alpha_L^2}{p_L\ln 2}$.
Nonetheless, a quadratic increase with $\alpha_L$ is also present.

\subsection{The thermodynamic limit}
We will first examine the previous quantities in
the ground and thermal state of (\ref{H1}) in the large $N$ limit, where we may
express all elements of $\rho_{L}$ in terms of the integrals (see appendix)
\begin{equation}
g_L=\frac{1}{\pi}\int_{0}^{\pi} \frac{\cos(L\omega)}{1+e^{\beta(B-J\cos\omega)}}\,
 d\omega\,,\label{gL}
\end{equation}
where $\beta=1/k_B T$ and $L=0,1,\ldots$, with $g_0=1/2+\langle s_z\rangle$ the
intensive magnetization.  We then obtain
\begin{eqnarray}
p^{\pm}_L&=&{\textstyle(g_0-\frac{1}{2}\pm\frac{1}{2})^2
-g_L^2}\,,\;\; p_L=g_0-g_0^2+g_L^2\,,\label{gg1}\\
\alpha_L&=&{\textstyle\frac{1}{2} {\rm Det}(A_L)},\,\,\,\,\,\,\, A_{ij}=
2 g_{i-j+1}-\delta_{i,j-1}\,,\label{gg2}
\end{eqnarray}
with $A_L$ the first $L\times L$ block of the matrix of elements $A_{ij}$
($i,j=1,\ldots,L$). Thus, $\alpha_1=g_1$, $\alpha_2=g_2(1-2g_0)+2g_1^2$, etc.

{\it Ground state results.} At $T=0$, all correlations vanish for $|B|>J$,
where the ground state is fully aligned along $z$ ($\alpha_L=0$, $p^+_L=1$
$\forall$ $L$). For $|B|<J$ we obtain instead
\begin{equation}
g_L=\frac{\sin(\omega L)}{L\pi}\,,\;\;
\omega=\arccos(B/J)\,,\label{ww}
\end{equation}
with $g_0=\omega/\pi$.

\begin{figure}
\centerline{\scalebox{.65}{\includegraphics{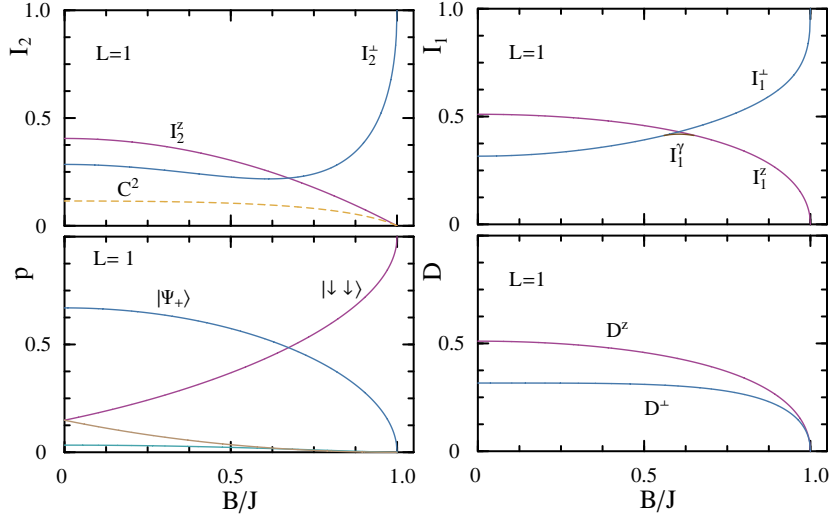}}} \vspace*{-0cm}
\caption{(Color online) Results for the geometric discord $I_2$ (top left), the
information deficit $I_1$ (top right), the quantum discord $D$ (bottom right)
and the eigenvalues of the reduced density matrix $\rho_{L}$ (bottom left) for
a pair of contiguous spins ($L=1$) in the ground state of the $XX$ chain in the
thermodynamic limit, as a function of the scaled transverse field. Superscripts
$z$ and $\perp$ denote the results for spin measurements parallel and
perpendicular to the field. In the top right panel the intermediate minimum
$I_1^\gamma$ in the small crossover region is also shown. The dashed line in
the top left panel depicts the square of the concurrence $C$. The minimum
$I_\nu$ ($\nu=1,2$) corresponds to $I_\nu^{\perp}$ essentially in the region
where the dominant eigenvector of $\rho_{L}$ is the Bell state
$|\Psi_+\rangle$.}
 \label{f1}
\end{figure}

\begin{figure}

\centerline{\scalebox{.65}{\includegraphics{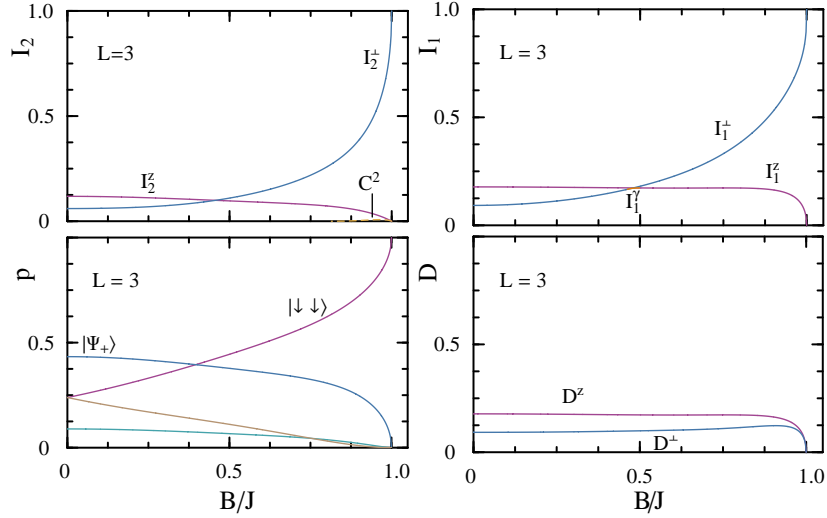}}} \vspace*{-0cm}
\caption{(Color online) The same quantities of Fig.\ \ref{f1} for third
neighbors ($L=3$).}
 \label{f2}
\end{figure}

Results for $I_2$, $I_1$, $D$ and the eigenvalues of $\rho_{L}$ are shown in
Figs.\ \ref{f1}--\ref{f2} for $L=1$ and $3$. It is first verified that while
the minimum quantum discord corresponds to $D^{\perp}$ $\forall$ $|B|<J$, the
minimum geometric discord $I_2$ exhibits, for decreasing $B$, the expected
sharp $I_2^z\rightarrow I_2^{\perp}$ transition.
Moreover, for $L=1$, this transition takes place {\it exactly at the point
where the Bell state $|\Psi_+\rangle$ becomes dominant} in $\rho_{L}$, i.e.,
$B_c^L=B_t^L$. Remarkably, for $L=1$ this exact coincidence occurs {\it at
both zero and finite temperature}, as follows from Eqs.\
(\ref{gg1})--(\ref{gg2}): In this case $\alpha_1=g_1$ and the crossing
condition (\ref{critl}), $\alpha_1=p_1^--p_1$, implies
\begin{equation}
g_1={\textstyle\frac{1}{2}}-g_0\,,\label{tlc}
\end{equation}
at this point. In such a case $p_1-p_1^+=p_1^--p_1=\alpha_1$, so that
$\alpha_1^c=\alpha_1^t$ (Eq.\ (\ref{I2x})) and hence $B_t^L=B_c^L$ for $L=1$.
At $T=0$ we have, explicitly,
\begin{equation}
\alpha_1={\textstyle\frac{\sin\omega}{\pi}=\frac{\sqrt{1-B^2/J^2}}{\pi}}
\,,\end{equation}
and this transition occurs at $B_t\approx 0.67 J$, i.e.,
$\sin\omega=\pi/2-\omega$, corresponding to an intensive magnetization $\langle
s_z\rangle\approx -0.235$. It is also seen that $I_2\geq C^2$ $\forall$ $B$,
i.e., the geometric discord remains larger than the corresponding entanglement
monotone.

The behavior of the information deficit $I_1$ is similar, except that the
previous  transition is smoothed through a small {\it crossover region}
$0.55\alt B/J\alt0.67$ where an {\it intermediate} measurement
($0<\gamma<\pi/2$) provides the actual minimum: As $B$ decreases, the
minimizing angle $\gamma$ increases smoothly from $0$ to $\pi/2$ in this
interval.

For higher separations, the behavior is similar except that values of $I_f$ and
$D$ are lower and the transition field $B_t^L$ is shifted towards lower fields,
in agreement with the decrease of the field $B_c^L$ where $|\Psi_+\rangle$
becomes dominant, as seen in Fig.\ \ref{f2} for $L=3$. $B_t^L$ remains close to
$B_c^L$ but the agreement is  not exact. The quantum discord continues to be
minimized by a perpendicular measurement $\forall$ $|B|<J$. Notice that in this
case the concurrence is very small and non-zero just in the vicinity of $B=J$,
whereas all $I_f$ and $D$ remain non-zero $\forall$ $|B|<J$, $\forall$ $L$.

Results for large separations $L\agt 3$ can be fully understood with the small
$\alpha_L,\delta$ expressions (\ref{Iza}), (\ref{grg}) and (\ref{Dperpa}).
For large $L$ we may neglect $g_L$ in $p_L^\pm$ and $p_L$, in which case
$p_L\approx\sqrt{p_L^+p_L^-}= \frac{\omega}{\pi}(1-\frac{\omega}{\pi})$, while
\begin{equation}\alpha_L=\eta_L/\sqrt{L}\,,\end{equation}
with $\eta_L$ approaching a {\it finite} value as $L$ increases
($\eta_L\rightarrow 0.294$ at $B=0$, decreasing with increasing $B$). For
sufficiently large $L$, Eq.\ (\ref{Iza}) then leads  to
\begin{equation}
I_f=I_f^z\approx k_f\eta_L^2/L\,,\;\;\; |B|>B_t^L\,,
 \end{equation}
with $k_f=|f''(p_L)|$ ($k_f=4$ in $I_2$ and $\frac{1}{p_L\ln 2}$ in $I_1$).
Hence, {\it all $I_f$'s decrease as $L^{-1}$ for increasing separations $L$}.

\begin{figure}

\centerline{\scalebox{.7}{\includegraphics{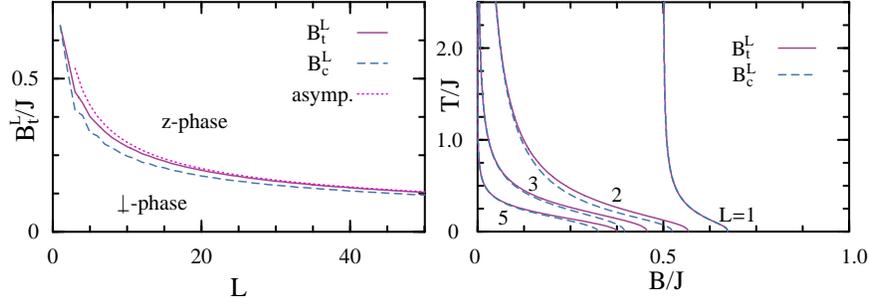}}} \vspace*{-0cm}
\caption{(Color online). Left: The $T=0$ transition field $B_t^L$ where the
measurement minimizing the geometric discord $I_2$ changes from perpendicular
to parallel, as a function of the separation $L$ (solid line), together with
the $T=0$ field $B_c^L$ where the dominant eigenvector of $\rho_L$ changes from
a Bell state to an aligned state (dashed line). Both fields coincide for $L=1$
and $L\rightarrow\infty$. The asymptotic result (\ref{bas}) for large $L$ is
also depicted (dotted line). Right: The transition fields $B_t^L(T)$ of the
geometric discord at finite temperatures, for $L=1,2,3$ and 5, such that
$I_2=I_2^{\perp}$ ($I_2^z$) for $B<B_t^L(T)$ ($>B_t^L(T))$. Dashed lines depict
again the fields $B_c^L(T)$ below which the Bell state is the dominant
eigenvector of $\rho_L$. For $L=1$, both fields coincide exactly $\forall$ $T$,
approaching $J/2$ for high $T$, whereas for $L\geq 2$ they merge for high $T$,
vanishing as $(J/T)^{L-1}$ (Eq.\ (\ref{Bctl})).}
 \label{f3}
\end{figure}

For large $L$ the transition field $B_t^L$  becomes small, so that for
$|B|<B_t^L$ we may employ the lower row of Eq.\ (\ref{grg}), with
$\delta\approx -B/(\pi J)$, since (\ref{ww})  implies here $\omega\approx
\pi/2-B/J$  and hence $p^\pm_L\approx\frac{1}{4}\mp B/(\pi J)$:
\begin{equation}
I_f=I_f^{\perp}\approx {\textstyle\frac{1}{2}}k_f [\eta^2_L/L+B^2/(\pi J)^2]\,,
\;\;\;|B|<B_t\label{Be2}\end{equation}
where  $k_f=|f''(1/4)|$  and
\begin{equation}
B_t^L\approx \pi\eta_L J/\sqrt{L}\,,\label{bas}
\end{equation}
as determined from the condition $I_f^{\perp}=I_f^z$ ($\eta_L\approx 0.294$ in
(\ref{Be2})--(\ref{bas})). This last equation coincides for large $L$ with the
condition $\alpha_L=p_L^- -p_L$ (Eq.\ (\ref{critl})), so that in this limit
$B_t^L=B_c^L$, as seen in the left panel of Fig.\ (\ref{f3}): The field
(\ref{bas}) also determines the onset as B decreases of $|\Psi^+\rangle$ as
dominant eigenstate of $\rho_L$. This field decreases then as  $L^{-1/2}$.

On the other hand, the quantum discord exhibits no transition: it
is verified that $D=D^{\perp}$ $\forall$ $B$, $L$. Its expression for large $L$
can be obtained directly from Eq.\ (\ref{Dperpa}) and implies $D\propto L^{-1}$
for large $L$, like $I_f$:
\begin{equation}D=D^\perp\approx k_D\,\eta_L^2/L,\label{DL}\end{equation}
where $k_D=\frac{1}{\ln 2}(\frac{1}{p_L}-\frac{{\rm arctanh}\,
2\delta}{\delta})$ with $\delta=\omega/\pi-1/2$. For  $B\rightarrow
0$, $\delta\rightarrow 0$ while $p_L\rightarrow 1/4$, and $D^\perp\rightarrow
I_1^\perp$.

We finally notice that for $B\rightarrow J$, Eq.\ (\ref{ww}) leads to
$\omega\approx \sqrt{2(1-B/J)}$, and hence $\alpha_L\approx g_L\approx
\omega/\pi$ $\forall$ $L$ at leading order. We then obtain the common
{$L$-independent} limits
\begin{eqnarray}
I_2&\approx& {8(1-B/J)/\pi^2\,,\;\; I_1\approx\sqrt{I_2}}\,,\;\;(B\rightarrow
 J)\label{bcj}\end{eqnarray}
with $D\approx I_1$ at leading order. The independence of separation for
$B\rightarrow J$ is verified and easily understood in the finite case (see next
section).

\begin{figure}

\centerline{\scalebox{.65}{\includegraphics{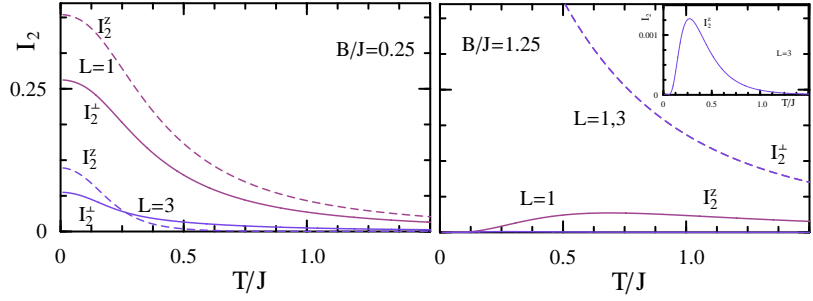}}} \vspace*{-0cm}
\caption{(Color online) The geometric discord $I_2$ vs.\ temperature at fixed
field, for first and third neighbors. At fields $B<B_c=J$ (left panel), $I_2$
decreases with increasing $T$ and a transition  $I_2^{\perp}\rightarrow I_2^z$
can take place, as seen here for $L=3$. For  $B>B_c$, $I_2=I_2^z$ first
increases at low $T$, although this revival becomes very small as $L$
increases, as seen in the inset for $L=3$. For high $T$, $I_2\propto
(T/J)^{-2L}$ (Eq.\ (\ref{IaT})).}
 \label{f4}
\end{figure}

{\it Finite temperatures.} As $T$ increases, $\alpha_L$  decreases for fields
$|B|<J$ (actually $|B|<J-\varepsilon_L$, with $\varepsilon_L$ small), implying
the decrease of all quantum correlations in this region.  Nonetheless, while
the concurrence (and hence entanglement) terminates at a finite $T$
\cite{CR.07}, the quantum discord and all $I_f$'s vanish only asymptotically
for high $T$. In addition, for $T>0$ a small but finite value of $D$ and $I_f$
will also arise for $B>J$ (Fig.\ \ref{f4}), as correlated excited states become
accessible.

Setting $k_B=1$, at high temperatures $T\gg {\rm Max}[J,B]$ Eq.\
(\ref{gg1})--(\ref{gg2})  lead to
\[{\textstyle g_0\approx \frac{1}{2}-\frac{B}{4T}\,,
\;\;g_1\approx \frac{J}{8T}}\,,\]
with $g_L=O(T^{-3})$ or higher for $L\geq 2$. Hence, in this limit we obtain,
at leading non-zero order,
\begin{equation}
p^\pm_L\approx{\textstyle\frac{1}{4}(1\mp
B/T)\,,\;\;p_L\approx\frac{1}{4}}\,,\;\;
\alpha_L\approx{\textstyle\frac{1}{2}}(J/4T)^{L}\,,\label{at}\end{equation}
implying that we may directly apply Eqs.\ (\ref{grg}) and (\ref{Dperpa}).
Therefore, $I_f$ and $D$ will vanish {\it exponentially} with increasing $L$,
i.e., $I_f,D\propto (T/J)^{-2L}$. Nonetheless, for all $I_f$'s there is still a
transition field $B_t^L$ $\forall$ $T$ such that $I_f^{\perp}<I_f^z$ for
$|B|<B_t^L$, with $B_t^L$ decreasing with increasing $T$ and approaching the
field $B_c^L$ for the onset of $|\Psi_+\rangle$ as the dominant eigenstate of
$\rho_L$. The final result for high $T$ derived from Eq.\ (\ref{grg}) is
\begin{equation}
I_f=\left\{\begin{array}{lr}I_f^z\approx
 \frac{k_f}{4}(\frac{J}{4T})^{2L}\,,&\;|B|>B_t^L\\
I_f^\perp\approx \frac{k_f}{2}(\frac{1}{4}
(\frac{J}{4T})^{2L}+\frac{B^2}{(4T)^2})\,,&\;|B|<B_t^L
\end{array}\right.\,,\label{IaT}\end{equation}
where $k_f=|f''(p_L)|\approx |f''(1/4)|$ and
\begin{equation}
 B_t^L\approx {\frac{J}{2}(\frac{J}{4T})^{L-1}}\label{Bctl}\,,
 \end{equation}
as determined from the condition $I_f^\perp=I_f^z$, which coincides in this
limit with that derived from the crossing condition (\ref{critl}). Hence, for
first neighbors ($L=1$) $B_t^L$ approaches for high $T$ the {\it finite} limit
$J/2$, whereas for $L\geq 2$  it decreases as $(J/T)^{L-1}$, as verified in the
right panel of Fig.\ \ref{f3} for $I_2$. In this limit the transition fields
$B_t^L$ approach  $B_c^L$ $\forall$ $I_f$. For lower $T$ they remain quite
close. It is also seen in Fig.\ \ref{f3} that in the case of $I_2$,
$B_t^L=B_c^L$ $\forall$ $T$ for $L=1$, as previously demonstrated.

In contrast $D=D^\perp$ $\forall$ $B,T$, with (Eq.\ (\ref{Dperpa}))
\begin{equation}
D^\perp\approx {\textstyle\frac{k_D}{4}(\frac{J}{4T})^{2L}}\,,
\end{equation}
for high $T$, where $k_D\approx\frac{2}{\ln 2}$. Again, $D^\perp\approx
I_1^\perp$ for $B\rightarrow 0$.

We finally notice that for $T>0$ and strong fields $B\gg J,T$, we have
\[g_L\approx \frac{e^{-\beta B}}{\pi}\int_{0}^\pi
 e^{\beta J \cos\omega}\cos(L\omega)d\omega=e^{-\beta B}I_L(\beta J)\,,\]
where $I_L(x)$ denotes the modified Bessel function of the first kind
($I_L(x)\approx e^x/\sqrt{2\pi x}$ for $x\rightarrow \infty$ while
$I_L(x)\approx (x/2)^L/L!$ for $x\rightarrow 0$). Hence, in this limit $g_L$
decreases exponentially with the field, with $p_L\approx g_0$ and
$\alpha_L\approx g_L$. The geometric discord then becomes
\begin{equation} I_2\approx 4e^{-2B/T}I_L^2(J/T)\,,\end{equation}
decreasing as $e^{-2B/T}$ for strong fields and also quite fast with separation
if $B\gg T\gg J$ ($I_L(J/T)\approx (J/2T)^L/L!$). On the other hand,  $I_1$ and
$D$ will decrease for strong fields as $\alpha_L$ ($\propto e^{-B/T}$).

\subsection{The finite case}
In a finite chain, the exact ground state has a definite discrete
magnetization $M$. Therefore, it will exhibit $N$ transitions
$M\rightarrow M+1$ as the field decreases from above $B_c=J$,
starting at $M=-N/2$ for $B>B_c$. In the cyclic case the critical
fields are given  by \cite{CR.07}
\begin{equation}
B_k=J{\textstyle\frac{\cos[\pi(k-1/2)/N]}{\cos[\pi/(2N)]}}
\,,\;k=1,\ldots,N\label{BCM}
\end{equation}
such that $M=k-N/2$ for $B_{k+1}<B<B_k$, with $B_1=J$, $B_N=-J$.
For $N\rightarrow\infty$ Eq. (\ref{BCM}) reduces to Eq.\
(\ref{ww}) ($B=J\cos\omega$, with $\omega/\pi=k/N=1/2+M/N$).
Details of the exact calculation for the finite case at $0$ and
finite $T$ are given in the Appendix.

\begin{figure}
\centerline{\scalebox{.8}{\includegraphics{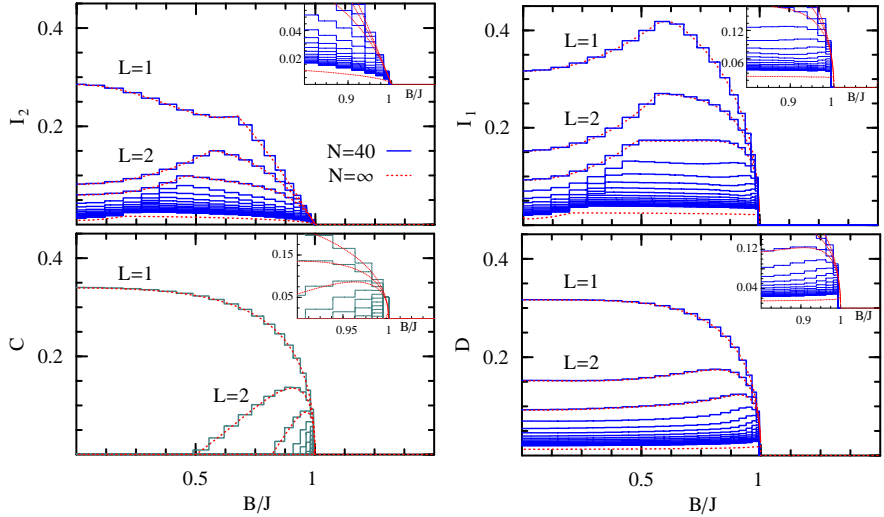}}}

\vspace*{-0cm}
\caption{(Color online) The minimum geometric
discord $I_2$ (top left), information deficit $I_1$ (top right)
and quantum discord $D$ (bottom right) for spin pairs with
separation $L=1,\ldots,N/2$ in the  ground state of a finite cyclic chain
of $N=40$ spins as a function of the scaled transverse magnetic
field. For reference the concurrence (bottom left) is also
depicted. The dotted lines depict the thermodynamic limit for
separations $L=1,2,3$ and $N/2$. In each panel the inset depicts
the vicinity of the critical field $B_c=J$, where all curves reach
a common value for all separations $L$ (Eqs.\
(\ref{I2w})--(\ref{Dw})).}
 \label{f5}
\end{figure}

Accordingly, all measures $I_f$ and $D$ will exhibit at $T=0$ a stepwise
behavior, starting from $0$ for $B>B_c$, which can be appreciated in Fig.\
\ref{f5} and which is centered around the result for the thermodynamic limit
(also depicted for $L=1,2,3$ and $N/2$) for  $L\alt N/4$. Just for large $L\agt
N/4$, the finite result becomes larger. In contrast, the concurrence is
non-zero for large $L$ just in the immediate vicinity of $B_c=J$.

Actually, as shown in the insets of Fig.\ 5, all measures $I_f$,  $D$ and $C$
acquire a {\it common value for all separations $L$} for $B\rightarrow J$,
i.e., in the first interval $B_2<B<B_1$, where $M=-N/2+1$ and the ground state
is the $W$-state
\[|\Psi_0\rangle=\frac{1}{\sqrt{N}}
(|\!\uparrow\downarrow\downarrow\ldots\rangle+\ldots+|
 \ldots\downarrow\downarrow\uparrow\rangle)\,.\]
This state leads to a $L$-independent rank 2 reduced state
$\rho_L$, with $p_L^+=0$, $p_L^-=1-2/N$ and $p_L=\alpha_L=1/N$ in
(\ref{rij}). For such state  we obtain, if $N\geq 4$,
\begin{equation}
I_2=I_2^z={\textstyle\frac{4}{N^2}}=C^2,\;\;I_1=I_1^z
 =\sqrt{I_2}\,,\label{I2w}
 \end{equation}
in agreement with the thermodynamic limit result (\ref{bcj}) (for large $N$ the
second critical field is $B_2\approx J(1-\frac{\pi^2}{N^2})$ for large $N$ and
hence, $\frac{8}{\pi^2}(1-\frac{B}{J})\approx \frac{4}{N^2}$ if
$B=\frac{B_1+B_2}{2}$).  Note that for this state, $\alpha_L\leq
\alpha_c=p_L^--p_L$ $\forall$ $N\geq 4$ (just for $N=3$, where
$\alpha_L>\alpha_c$, a perpendicular measurement is preferred in both $I_2$ and
$I_1$). In contrast, $D$  is minimized by a perpendicular measurement $\forall$
$N$, with
\begin{equation}
D^\perp\approx {\textstyle\frac{2}{N}-\frac{1}{N^2}\log_2 (N/e)}\,,
\label{Dw}\end{equation}
for large $N$ (though $D^{\perp}\approx D^z=I_1^z$ at leading order).

For lower fields, it is seen that for small $L\geq 2$, $I_2$ is maximum at the
parallel-perpendicular transition. Such maximum becomes flattened in $I_1$ and
is absent in the quantum discord $D$, since the latter is minimized by a
perpendicular measurement $\forall$ $B<J$ and $L$. For $L>1$ its maximum is
attained in the vicinity of $B_c=J$.

\begin{figure}
\centerline{\scalebox{.7}{\includegraphics{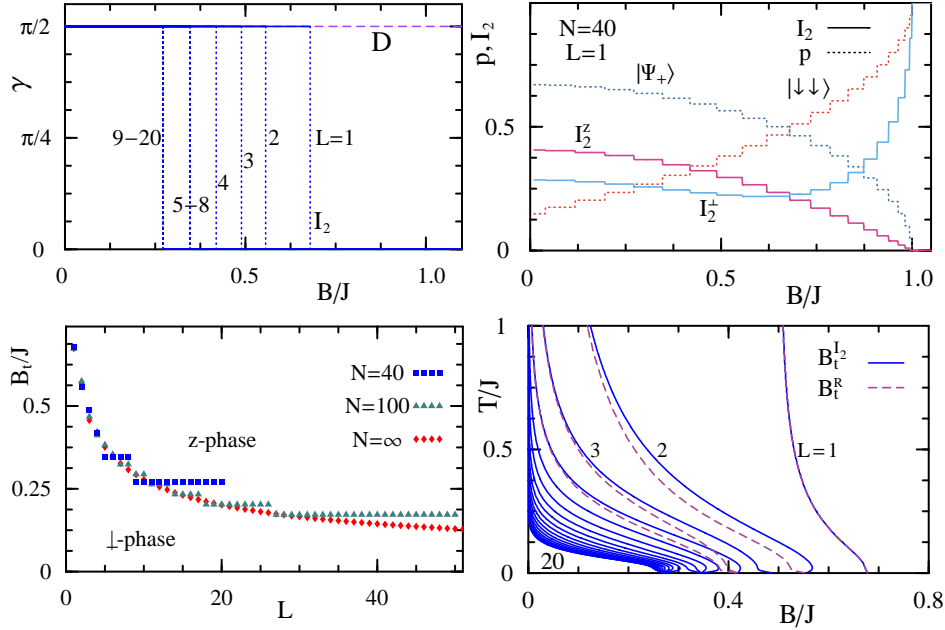}}}
\vspace*{-0cm}
\caption{(Color online). Top: Left: The minimizing angle for the geometric
discord $I_2$ as a function of the magnetic field for spin pairs with
separations $L=1,\ldots,N/2$, in the finite chain of Fig.\ \ref{f5}. Dotted
lines indicate the sharp $\perp\rightarrow z$ transitions for different  $L$.
No transition occurs in the quantum discord $D$ (dashed line), where
$\gamma=\pi/2$ $\forall$ $B$ and $L$. Right: Results for the geometric discord
$I_2^{\perp}$ and $I_2^z$ (solid lines) for $N=40$ and $L=1$, together with the
two dominant eigenvalues of $\rho_1$ (dotted lines). Both cross at the same
step. Bottom: Left: Exact transition fields $B_t^L$  delimiting the $\perp$ and
$z$ phases of $I_2$ at $T=0$  for  $N=40$, $N=100$ and the thermodynamic limit.
Right: The geometric discord ``phase'' diagram in the finite chain of $N=40$
spins, for all separations $L=1,\ldots,N/2$ (solid lines). The $z$ ($\perp$)
phases lie to the right (left) of these curves. Dashed lines depict the fields
$B_c^L(T)$ for $L\leq 4$, below which the Bell state becomes dominant in
$\rho_L$.  }
 \label{f6}
\end{figure}

The minimizing angles for $I_2$ and $D$ in the finite case of Fig.\ \ref{f5}
are depicted in Fig.\ \ref{f6}. For $I_2$, it exhibits  the sharp transition
from $\gamma=0$ ($z$ phase) to $\gamma=\pi/2$ ($\perp$ phase)  as $B$
decreases, which now occurs {\it at one of transition fields} (\ref{BCM})
($B_t^L=B_{k}$ for some $L$-dependent $k$). For $L=1$ the measurement
transition signals exactly that ground state transition $M\rightarrow M+1$
where $\rho_L$ changes its dominant eigenstate, as clearly depicted in the top
right panel of Fig.\ \ref{f6} (where it corresponds to $k=11$ in (\ref{BCM})),
while for larger $L$ both transitions are very close. As seen in the left
panels of Fig.\ \ref{f6}, as $L$ increases the transition fields for different
$L$ begin to merge, coinciding for large $L\agt N/4$, while as $N$ increases
they approach the thermodynamic limit result for $L\alt N/4$, becoming then
constant. A similar picture  is obtained for the minimizing angle of $I_1$,
although in this case the measurement transition can occur in two or three
``steps'', reminiscent of the smoothed transition of the thermodynamic limit.

The bottom right panel in  Fig.\ \ref{f6} depicts the finite temperature
geometric discord ``phase'' diagram according to the minimizing measurement for
$N=40$ spins (fields $B_t^L$), together with the fields $B_c^L$ where dominant
eigenstate changes from the Bell state to an aligned state,  for all
separations $L=1,\ldots,N/2$. For $L=1$ there is again almost exact coincidence
between both fields for all $T$, since the deviation from the thermodynamic
limit condition (\ref{tlc}) is small. For larger $L$ the agreement is not exact
for low $T$, but they become again coincident for high temperatures $\forall$
$L$, where deviations from the thermodynamic limit results become small.

\section{Conclusions}
We have examined the behavior of discord-type measures of quantum correlations
for the case of spin pairs in the cyclic XX chain. Their behavior is
substantially different from that of the pair entanglement, acquiring at $T=0$
non-zero values for all pair separations $L$ if $B<B_c$ and decaying only as
$L^{-1}$ for large $L$. Moreover, they remain non-zero for all temperatures,
decaying as $T^{-2L}$ for sufficiently high $T$. Thus, they all exhibit the
same ``universal'' asymptotics, independently of the particular choice of
entropic function in $I_f$. It can then be most easily accessed through the
geometric discord, which offers the simplest evaluation. The ensuing picture
is, consequently, quite different from that exhibited by the pair entanglement
\cite{CR.07}, which, although reaching full range in the immediate vicinity of
$B_c$, is appreciable just for the first few neighbors, as seen in Fig.\
\ref{f5}, and strictly vanishes beyond a low limit temperature. Hence, critical
systems like the XX chain seem to offer vast possibilities for discord-type
quantum correlations between close or distant pairs.

The second important result is that in spite of the similar behavior, these
measures exhibit substantial differences in  the minimizing local spin
measurement that defines them. The quantum discord, which minimizes a
conditional entropy, always prefers here measurements along a direction
orthogonal to the transverse field, even if correlations are weak (i.e., large
$L$, high $T$ or strong fields $B$ if $T>0$). The information deficit-type
measures, which evaluate the minimum global information loss due to such
measurement and include the geometric discord and the one-way information
deficit, exhibit instead a transition in the optimum measurement, from
perpendicular to parallel to the field as the latter increases, present for all
pair separations and at all temperatures. Such difference was previously
observed in certain two-qubit and two-qutrit states \cite{RCC.11,RMC.12}.

In the present model such behavior is a signature of the transition exhibited by the
dominant eigenstate of the reduced state of the pair, which changes from a
maximally entangled state to a separable state in the immediate
vicinity of the measurement transition. Hence, the latter reveals an actual relevant
change in the structure of the reduced state.  Moreover, for contiguous pairs
and in the case of the geometric discord, both transitions occur {\it exactly}
at the same field, at all temperatures. For general separations there is
also exact agreement between both fields at high $T$, for {\it all measures $I_f$}.
In the finite chain the $T=0$ measurement transition coincides of course with
one of the ground state magnetization transitions $M\rightarrow M+1$. These results
indicate that the ``least disturbing'' local measurement optimizing these quantities
can  be significantly different from that minimizing the quantum discord, even though they coincide exactly in some regimes, being essentially affected by the main component of the reduced state. Its changes can then be used to characterize different quantum regimes, even when entanglement is absent.

The authors acknowledge support from CONICET (LC, NC) and CIC
(RR) of Argentina.

\appendix
\section{Exact solution of the cyclic chain }
We give here a brief summary of the method employed for obtaining the exact
solution of the cyclic $XX$ chain for both finite $N$ and the thermodynamic
limit, at both $0$ and finite $T$ \cite{CR.07}. Through the Jordan-Wigner
transformation \cite{LSM.61}, and for each value $\sigma=\pm 1$ of the $S_z$
parity $P_z=\exp[i\pi (S_z+N/2)]$, the $XX$ Hamiltonian can be mapped exactly
to the fermionic Hamiltonian
\begin{eqnarray}
H_{\sigma}&=&{\textstyle\sum_{j=1}^N B(c^\dagger_jc_j-\frac{1}{2})-\frac{1}{2}
J(1-\delta_{jN}\delta_{\sigma 1}) (c^\dagger_jc_{j+1}+c^\dagger_{j+1}c_j)}
 \label{Hf1}\end{eqnarray}
where $N+1\equiv 1$ and $c_j, c^\dagger_j$ denote fermion annihilation and
creation operators. Eq.\ (\ref{Hf1}) can be solved exactly through a discrete
Fourier transform $c^\dagger_j={\frac{1}{\sqrt{N}}} \sum_{k\in
K_\sigma}e^{i\omega_k j}c'^\dagger_{k}$ to fermion operators $c'_k$, where
$\omega_{k}=2\pi k/N$ and $k$ is {\it half-integer (integer)} for $\sigma=1$
($-1$), i.e.,
$K_{\sigma}=\{-[\frac{N}{2}]+\delta_{\sigma},\ldots,
[\frac{N-1}{2}]+\delta_{\sigma}\}$
with $[\ldots]$ the integer part and $\delta_1=\frac{1}{2}$, $\delta_{-1}=0$.
We then obtain
\begin{equation}
H_\sigma=\sum_{k\in K_\sigma}\lambda_k
(c'^\dagger_{k}c'_{k}-{\textstyle\frac{1}{2}}),
\;\;\;\;\;\lambda_k=b-v\cos\omega_k\,.\label{lk}
 \end{equation}
The $2^N$ energies are then $\sum_{k\in K_\sigma}\lambda_k(N_k-1/2)$, where
$N_k=0,1$ and $\sigma=(-1)^{\sum_k N_k}$. The single fermion energies
$\lambda_k$ depend on the global parity $\sigma$ and are degenerate
($\lambda_k=\lambda_{-k}$) for $|k|\neq 0,N/2$. A careful comparison of the
ensuing levels leads to the critical fields (\ref{BCM}).

The exact partition function $Z$ of the spin system corresponds to the full
grand-canonical ensemble of the fermionic representation. However, due to the
parity dependence of the latter, this requires a (fermion) {\it  number parity
projected statistics} \cite{CR.07}. $Z$ can then be written as a sum of
partition functions for each parity,
\begin{equation} Z=
{\rm Tr}\!\sum_{\sigma=\pm 1}P_\sigma e^{-\beta H_{\sigma}}=
{\textstyle\frac{1}{2}}\! \sum_{\sigma=\pm 1} (Z^\sigma_0+\sigma Z^\sigma_1)\,,
 \label{Zp}\end{equation}
where $P_\sigma=\frac{1}{2}(1+\sigma P_z)$ is the projector onto parity
$\sigma$ and $Z_\nu^\sigma=e^{\beta BN/2}\prod_{k\in K_\sigma}(1+(-1)^\nu
e^{-\beta\lambda_k})$ for $\nu=0,1$. The thermal average of an operator $O$ can
then be written as
\begin{eqnarray}
\langle O\rangle&=&{\textstyle\frac{1}{2}} Z^{-1}\sum_{\sigma=\pm 1}
(Z^\sigma_0\langle O\rangle_0^\sigma+\sigma Z^\sigma_1\langle
 O\rangle_1^\sigma)\,,\label{Om}\end{eqnarray}
where $\langle O\rangle_{\nu}^{\sigma}=(Z_\nu^\sigma)^{-1} {\rm Tr}\,[P_z^\nu
e^{-\beta H_{\sigma}}O]$. For many-body fermion operators $O$, the thermal
version of Wick's theorem {\it cannot} be applied in the final average
(\ref{Om}), but it {\it can} be applied for evaluating the partial averages
$\langle O\rangle^\sigma_\nu$, in terms of the basic contractions ($L=|i-j|$) 
\begin{equation}
g_L\equiv\langle {c}^\dagger_i c_{j}\rangle_\nu^{\sigma}=N^{-1}\sum_{k\in
K_{\sigma}} \langle c'^\dagger_kc'_k\rangle^{\sigma}_{\nu}\cos(L\omega_k)\,,
\end{equation}
where $\langle {c'}^\dagger_k c'_{k}\rangle_\nu^{\sigma}=[1+(-1)^\nu
e^{\beta\lambda_k}]^{-1}$. As $s_{iz}=c^\dagger_ic_i-\frac{1}{2}$,
this leads to $\langle s_{iz}\rangle^\nu_\sigma=g_0-\frac{1}{2}$ and
\[
{\textstyle\langle(s_{iz}+\frac{1}{2})(s_{jz}+\frac{1}{2})\rangle_\nu^\sigma}
=g_0^2-g_L^2\,\,,\;\;
\langle s_{i+}s_{j-}\rangle_\nu^\sigma={\textstyle\frac{1}{2}}{\rm Det}(A_L)
\]
where $s_{j\pm}=s_{jx}\pm is_{jy}$ and $A_L$ is the $L\times L$ matrix of
elements $(A_L)_{ij}=2g_{i-j+1}-\delta_{i,j-1}$. All elements in (\ref{rij})
can then be analytically evaluated.

For  $N\rightarrow\infty$ and finite separations $L$, we can ignore parity
effects and directly employ Wick's theorem in terms of the final averages
$g_L=\langle c^\dagger_i c_j\rangle$, with sums over $k$ replaced by integrals
over $\omega\equiv \omega_k$. This leads to Eqs.\ (\ref{gL})--(\ref{gg2}). When
the ground state is non-degenerate, Eqs.\ (\ref{gg1})--(\ref{gg2}) can also be
applied for {\it finite} $N$ in the $T\rightarrow 0$ limit, using the exact
contractions $g_L\equiv \langle c^\dagger_ic_j\rangle_0= \frac{1}{N}\sum_{k\in
K_\sigma}N_k\cos(L\omega_k)$, with $N_k=0,1$ the occupation of level $k$.

\end{document}